\documentclass[twocolumn,superscriptaddress,showpacs,prl,amsmath,amssymb]{revtex4}
\usepackage{graphicx}
\usepackage{dcolumn}
\usepackage{bm}

\usepackage{color}

\begin{document}

\title{Are There Nuclear  Structure Effects on the Isoscalar Giant Monopole Resonance and  Nuclear Incompressibility  near $A\sim$ 90?}

\author{Y. K. Gupta}
\thanks{Permanent address: Nuclear Physics Division, Bhabha Atomic Research Centre, Mumbai, 400085, India} 
\affiliation{Physics Department, University of Notre Dame, Notre Dame, IN 46556, USA}

\author{U. Garg}
\affiliation{Physics Department, University of Notre Dame, Notre Dame, IN 46556, USA}
\author{K. B. Howard}
\affiliation{Physics Department, University of Notre Dame, Notre Dame, IN 46556, USA}
\author{J. T. Matta}
\thanks{Present address: Physics Division, Oak Ridge National Laboratory, Oak Ridge, TN 37380, USA}
\affiliation{Physics Department, University of Notre Dame, Notre Dame, IN 46556, USA}
\author{M. {\c S}enyi{\u g}it}
\thanks{Permanent address: Department of Physics, Faculty of Science, Ankara University, TR-06100 Tando{\u g}an, Ankara, Turkey}
\affiliation{Physics Department, University of Notre Dame, Notre Dame, IN 46556, USA}

\author{M. Itoh} 
\affiliation{Cyclotron and Radioisotope Center, Tohoku University, Sendai 980-8578, Japan}
\author{S. Ando} 
\affiliation{Cyclotron and Radioisotope Center, Tohoku University, Sendai 980-8578, Japan}
\author{T. Aoki} 
\affiliation{Cyclotron and Radioisotope Center, Tohoku University, Sendai 980-8578, Japan}
\author{A. Uchiyama} 
\affiliation{Cyclotron and Radioisotope Center, Tohoku University, Sendai 980-8578, Japan}

\author{S. Adachi}
\affiliation{Research Center for Nuclear Physics (RCNP), Osaka University, Osaka 567-0047, Japan}
\author{M. Fujiwara}
\affiliation{Research Center for Nuclear Physics (RCNP), Osaka University, Osaka 567-0047, Japan}
\author{C. Iwamoto}
\affiliation{Research Center for Nuclear Physics (RCNP), Osaka University, Osaka 567-0047, Japan}
\author{A. Tamii}
\affiliation{Research Center for Nuclear Physics (RCNP), Osaka University, Osaka 567-0047, Japan}

\author{H. Akimune}
\affiliation{Department of Physics, Konan University, Hyogo 658-8501, Japan}
\author{C. Kadono}
\affiliation{Department of Physics, Konan University, Hyogo 658-8501, Japan}
\author{Y. Matsuda}
\affiliation{Department of Physics, Konan University, Hyogo 658-8501, Japan}
\author{T. Nakahara}
\affiliation{Department of Physics, Konan University, Hyogo 658-8501, Japan}

\author{T. Furuno}
\affiliation{Department of Physics, Kyoto University, Kyoto 606-8502, Japan}
\author{T. Kawabata}
\affiliation{Department of Physics, Kyoto University, Kyoto 606-8502, Japan}
\author{M. Tsumura}
\affiliation{Department of Physics, Kyoto University, Kyoto 606-8502, Japan}

\author{M. N. Harakeh}
\affiliation{KVI-CART, University of Groningen, 9747 AA Groningen, The Netherlands}
\author{N.  Kalantar-Nayestanaki }
\affiliation{KVI-CART, University of Groningen, 9747 AA Groningen, The Netherlands}

\date{\today}

\begin{abstract}
``Background-free" spectra of inelastic $\alpha$-particle scattering have been measured  at a beam energy of 385 MeV in $^{90, 92}$Zr and $^{92}$Mo at extremely forward angles, including 0$^{\circ}$. 
The ISGMR strength distributions for the three nuclei coincide with each other, establishing clearly  that nuclear incompressibility is not influenced by nuclear shell structure near $A\sim$90 as was claimed in recent measurements.
\end{abstract}


\maketitle

Nuclear incompressibility is a fundamental quantity characterizing the equation of state (EOS) of nuclear matter  \cite{BohrMott}. A number of important phenomena such as the radii of neutron stars, the strength of supernova explosions, transverse flow in relativistic heavy-ion collisions, the nuclear skin thickness, etc. require a good understanding of the EOS of nuclear matter \cite{Harakeh_book, Lattimer2001}. The nuclear incompressibility for infinite nuclear matter, $K_{\infty }$,  may be determined experimentally from the compressional ``breathing mode" of nuclear density oscillation, the isoscalar giant monopole  resonance (ISGMR), in finite nuclei \cite{str1982, Treiner1981}. In the scaling model, the energy of the ISGMR is directly related to the nuclear incompressibility of the nucleus and is given by \cite{str1982}:
\begin{equation}
E_\mathrm{ISGMR}=\hbar\sqrt{\frac{K_{A}}{m\left<r^{2}\right>_{0}}},
\label{EISGMR}
\end{equation}

\noindent
where $K_{A}$ is  the  incompressibility of a nucleus with mass number $A$, $\left<r^{2}\right>_{0}$ is the ground state mean square 
radius, and  $m$ is the nucleon mass. The determination of $K_{\infty }$ from 
$K_{A}$ is achieved within a framework of self-consistent RPA calculations, using the widely accepted method described by Blaizot et al. \cite{Blaizot1995,Harakeh_book}. The presently accepted value of $K_{\infty }$, determined from ISGMR in ``standard" nuclei such as $^{90}$Zr and $^{208}$Pb, is 240 $\pm$ 20 MeV \cite{Colo_Giai_PRC2004, Todd-Rutel_PRL2005, colo_garg_sagawa,  pickarewickz2014}. 
Because the compressional modes are collective phenomena, the determination 
of $K_{\infty }$ should be independent of the choice of the nucleus, provided that approximately 100\% of the energy weighted sum rule (EWSR) fraction is exhausted in the ISGMR peak; this condition is satisfied for sufficiently heavy nuclei ($A\geq 90$) \cite{Harakeh_book}. The use of the aforementioned ``standard nuclei'' stems primarily from the relative ease in doing theoretical calculations for the doubly-magic nuclei.

In recent work by the Texas A \& M group \cite{YB_ZrMo2013, YB_Mo2015, Krishi_2015}, it has been claimed that the ISGMR strength distributions  vary in a rather dramatic manner in nuclei in the $A\sim$ 90 region. In particular, the $A$=92 nuclei, $^{92}$Zr and $^{92}$Mo,  emerged quite disparate from the 
others: The ISGMR energies  ($E_\mathrm{ISGMR}$) for $^{92}$Zr and $^{92}$Mo were observed to be, respectively, 1.22 and  2.80 MeV higher than that of $^{90}$Zr.  Consequently, the $K_{A}$ values determined for $^{92}$Zr and $^{92}$Mo were $\sim$27 MeV  and $\sim $56 MeV, respectively, higher than that of $^{90}$Zr. 
These results, if correct, imply significant nuclear structure contribution to the nuclear incompressibility in this mass region. Such nuclear structure effects have not been observed in any of the investigations of ISGMR going back to its first identification  in the late 1970's  \cite{Harakesh_prl1977, YBPRL1977} and, indeed, would be contrary to the standard hydrodynamical picture associated with this mode of collective oscillation \cite{LIPPARINI1989}. Furthermore, this would lead to a very uncertain determination of the nuclear-matter incompressibility, a key parameter of the equation of state (EOS) of dense nuclear matter that plays ``an important role in the supernova phenomenon, the structure of neutron stars, and in the mergers of compact objects (neutron stars and black holes)''\cite{lattpra}. It is also important to note that the only Òstructure effectsÓ observed in giant resonances so far are due to ground-state deformation. In addition, there is the apparent softness of nuclei away from the closed shells. Both these effects, entirely different from the claims made for the A$\sim$90 nuclei, are discussed later.

In this Letter, we  report on the ISGMR response in the $^{90, 92}$Zr and $^{92}$Mo nuclei from inelastic $\alpha$-scattering measurements at an energy of 385 MeV and free of all ``instrumental background". These nuclei are found to have virtually identical ISGMR responses, revealing no influence of open and/or closed shells for protons and/or neutrons on the ISGMR and, hence, on the nuclear incompressibility.
\begin{figure}[t] 
\centering\includegraphics [trim= 0.11mm 0.5mm 0.1mm 0.1mm,
angle=360, clip, height=0.21\textheight]{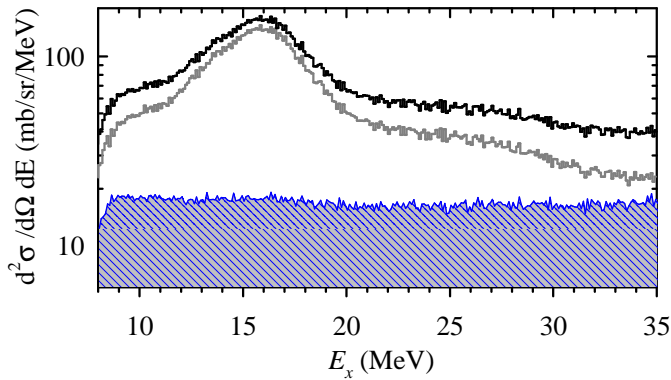}
\caption{Excitation-energy spectra for the $^{92}$Zr($\alpha, \alpha'$) reaction at $E_{\alpha}$=385 MeV at an
averaged spectrometer angle of $\theta_\mathrm{avg}$ = 0.7$^{\circ}$. The gray hatched region represents the instrumental background.
The solid black and  gray histograms show the energy spectra before and after the instrumental-background subtraction, respectively.}
\label{ZeroDeg}
\end{figure}

Inelastic scattering of 385-MeV $\alpha $ particles
was measured at the ring cyclotron facility
of the Research Center for Nuclear Physics (RCNP), Osaka
University. Self-supporting  foils of highly enriched  targets ($97.70\%$, $95.13\%$, and $97.37\%$ for $^{90}$Zr, $^{92}$Zr, and $^{92}$Mo, respectively)   were  used, with thicknesses ranging from 4.0 to 5.38 mg/cm$^{2}$. Inelastically scattered $\alpha$ particles were momentum analyzed with the high-resolution magnetic spectrometer ``Grand
Raiden" \cite{Fujiwara_GR}, and their  horizontal and vertical  positions were measured with a focal-plane detector
system composed of two position-sensitive multiwire drift
chambers (MWDCs) and two plastic scintillators \cite{Itoh_prc2003}. These detectors enabled particle identification and
reconstruction of the trajectories of scattered particles.  The
vertical-position spectrum obtained in the double-focusing
mode of the spectrometer was exploited to eliminate the
instrumental background \cite{Uchida_PLB2003,Itoh_prc2003}.
Fig.~\ref{ZeroDeg} shows the typical instrumental background, and excitation-energy spectra  before and after the background subtraction, for $^{92}$Zr as measured at an average spectrometer angle  $\theta_\mathrm{avg}$ = 0.7$^{\circ}$.

Data for elastic scattering and inelastic scattering to 2$^{+}$ and 3$^{-}$ states  for each nucleus were taken in the angular range of 5.0$^{\circ }$ to 26.5$^{\circ }$. Giant-resonance measurements  were performed  at very forward central angles of the spectrometer (from 0$^{\circ }$ to 9.5$^{\circ }$) and at magnetic-field settings corresponding to excitation energies in the range $E_x \sim$9.5--35.5 MeV.  Using the ray-tracing technique, the angular width  of 1.6$^{\circ }$ for each central angle  was divided into four equal regions during the offline data analysis. Data were also taken with a $^{12}$C target at each setting, providing a precise energy calibration. Energy losses in the target foils for the incident beam and outgoing $\alpha$ particles were taken into account.   

\begin{figure}[t] 
\centering\includegraphics [trim= 0.11mm 0.5mm 0.1mm 0.1mm,
angle=360, clip, height=0.32\textheight]{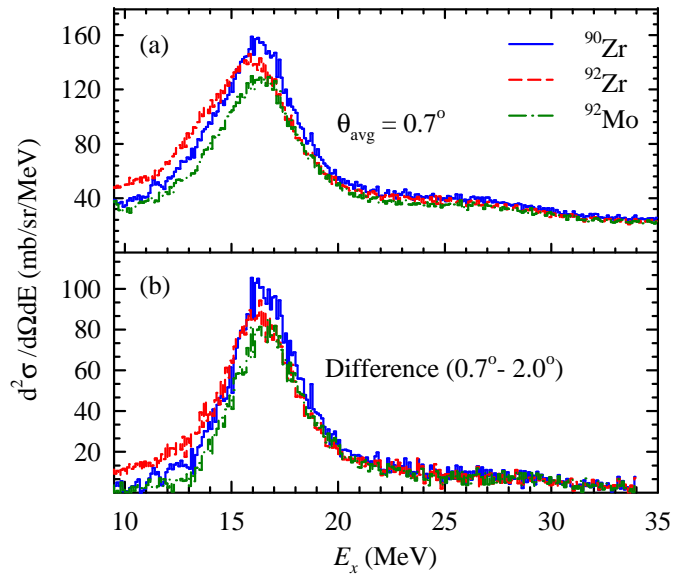}
\caption{ (Color online) (a) Excitation-energy spectra measured at an averaged angle 0.7$^{\circ}$ for the three nuclei, $^{90}$Zr (solid blue), $^{92}$Zr (red dash), and $^{92}$Mo (green dash-dot). (b) Difference spectra of averaged angles 0.7$^{\circ}$ and 2.0$^{\circ}$ for the same nuclei. The difference spectra comprise essentially  the monopole strength (see text).}
\label{ZeroDeg_DiffSpect}
\end{figure}

The excitation-energy spectra at $\theta_\mathrm{avg}$ = 0.7$^{\circ}$ for the three nuclei are overlaid in Fig.~\ref {ZeroDeg_DiffSpect}(a). The spectra near 0$^{\circ}$ scattering angle exhibit predominantly the monopole strength, and the 0.7$^{\circ}$ spectra for the three nuclei, shown in Fig.~\ref {ZeroDeg_DiffSpect}(a), are very similar; in particular, for excitation energies beyond 20 MeV, these are nearly identical whereas the results in Refs. \cite{YB_ZrMo2013, YB_Mo2015, Krishi_2015} had shown marked differences in this excitation-energy region. The minor differences in the low-energy part of the spectra  (below 16 MeV) are partly due to the different shapes of the ISGMR at low energy (see Fig. 4) 
and could also be partly due to the different contributions from 
the toroidal mode, which is not completely understood so far \cite{yogesh2}.
\begin{figure}[t]
\centering\includegraphics [trim= 0.5mm 0.5mm 0.1mm 0.1mm,
angle=360, clip, height=0.235\textheight]{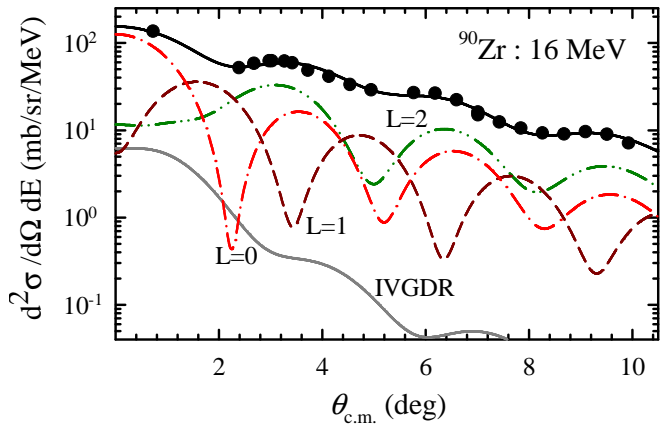}
\caption{(Color online) Typical angular distributions for inelastic $\alpha$
scattering from $^{90}$Zr at an excitation energy of 16 MeV. The solid line (black) through the data shows the sum of various multipole
components obtained from MDA. The dash-dotted (red), dashed (brown), and dash-double-dotted (green) curves
indicate contributions from $L$ = 0, 1, and 2, respectively, with the transferred angular momentum $L$  indicated for each of the curves. The solid gray line shows the IVGDR component.}
\label{MDA_90Zr}
\end{figure}

It has been recognized for quite some time now (see, e.g., Ref. \cite{BRANDENBURG1987}) that the ``difference-spectrum", obtained from subtracting the inelastic spectrum at the first minimum of the expected ISGMR angular distribution from that at 0$^{\circ}$ (where the ISGMR strength is maximal), essentially represents only the ISGMR strength. This is a consequence of the fact that all other multipolarities have relatively flat distributions in this angular region and, thus, are subtracted out in the ``difference-spectrum". In the present work, the difference spectra for averaged angles of 0.7$^{\circ}$ (maximal ISGMR strength) and 2.0$^{\circ}$ (first minimum of ISGMR strength) for all the three nuclei are also almost identical, as shown in  Fig.~\ref {ZeroDeg_DiffSpect}(b), again indicating similar ISGMR response in the three nuclei. In particular, the difference spectra beyond $E_x$=20 MeV fully coincide with each other, whereas the results in Refs. \cite{YB_ZrMo2013, YB_Mo2015, Krishi_2015} had shown dramatically different ISGMR strengths in this region.

In order to extract quantitative strengths for different multipolarities, we have employed the standard multipole-decomposition analysis (MDA) procedure \cite{Li_2010, YKG_PLB2015}. Experimental cross-sections were binned into 1-MeV intervals. The laboratory angular distributions for each excitation-energy bin were converted to the center-of-mass frame using the standard Jacobian and relativistic kinematics. A typical angular distribution for $^{90}$Zr 
at an excitation energy of 16 MeV is presented in Fig.~\ref{MDA_90Zr}.  The experimental double-differential cross sections are
expressed as linear combinations of calculated double-differential cross sections associated with different
multipoles as follows \cite{Li_2010, YKG_PLB2015}:
\begin{equation}
\frac{d^{2}\sigma ^{\mathrm{exp}} (\theta_{\mathrm{c.m.}}, E_{x})}{d\Omega dE}= \sum\limits_{L=0}^6 a_{L}(E_{x})\frac{d^{2}\sigma_{L}^{\mathrm{DWBA}}(\theta_{\mathrm{c.m.}}, E_{x}) }{d\Omega dE} 
\end{equation}
\noindent
where $a_{L}(E_{x})$ is EWSR fraction for the $L^{th}$ component and $\frac{d^{2}\sigma_{L}^{\mathrm{DWBA}} }{d\Omega dE} (\theta_{\mathrm{c.m.}}, E_{x})$ is the calculated DWBA cross section corresponding to 100\% EWSR for the $L^{th}$ multipole at excitation energy $E_{x}$. The isovector giant dipole resonance (IVGDR) contribution
was subtracted out of the experimental data prior to the fitting procedure \cite{Darshana2012, Li_2010}. 
The DWBA calculations were performed by following the method of Satchler and Khoa \cite{Satchler_Khoa1997} using the density-dependent single-folding model, with a Gaussian $\alpha$-nucleon potential for the real part and a Woods-Saxon
imaginary term \cite{Li_2010}. We used transition densities and sum rules for various multipolarities as described in Refs. \cite{Harakeh_book,Satchler1987,Harakeh1981}. 
The EWSR fractions, $a_{L}(E_{x})$, are determined from a $\chi^{2}$ minimization. Although we employed DWBA calculated cross sections up to $L$=6 in the MDA, the strengths could be reliably obtained only up to $L$=3 due to the limited angular range. 
The MDA fit to the angular distribution data for the  energy bin at $E_{x}$=16 MeV in $^{90}$Zr, as well as the dominant $L$ contributions are presented in Fig.~\ref{MDA_90Zr}. 
\begin{figure}
\centering\includegraphics [trim= 0.11mm 0.5mm 0.1mm 0.1mm,angle=360, clip, width=0.37\textheight]{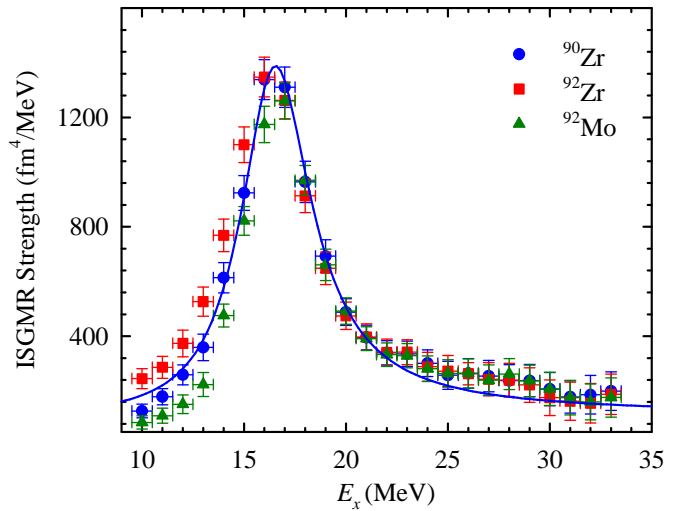}
\caption{(Color online) ISGMR  strength distributions for all the three nuclei, $^{90}$Zr (blue circles), $^{92}$Zr (red squares), and $^{92}$Mo (green triangles). The solid line represents the Lorentzian fit for $^{90}$Zr.}
\label{ISGMR}
\end{figure}

The optical model parameters (OMPs) used in the DWBA calculations were determined for each nucleus from elastic scattering angular distributions.  
The imaginary potential parameters ($W$, $R_{I}$, and $a_{I}$), together with the depth of the real part, $V$, were obtained  by fitting the elastic-scattering cross sections using the computer code PTOLEMY \cite{ptolemy1,ptolemy2}. Using the $B(E2)$ and $B(E3)$ transition probabilities from the literature, and the OMPs thus obtained, the angular distributions for the 2$^{+}$ and 3$^{-}$ states for each nucleus were calculated within the  same DWBA framework. Good agreement between the calculated and experimental angular distributions  for the 2$^{+}$ and 3$^{-}$ states for each nucleus established the appropriateness of the OMPs \cite{yogesh2}. 

Experimentally determined ISGMR strength distributions are displayed in Fig.~\ref{ISGMR} for the three nuclei investigated in the present work. The  strength distributions, each consisting of a single, broad peak at $E_{x}\sim$16.5 MeV, coincide with each other within experimental uncertainties, with small differences at the low-energy side.  
Again, the ISGMR strength in the high-excitation-energy range for $^{92}$Zr and $^{92}$Mo  is identical to that in $^{90}$Zr; the results in Refs. \cite{YB_ZrMo2013, YB_Mo2015, Krishi_2015} had shown  marked deviations instead, leading to quite different $E_\mathrm{ISGMR}$ values. We also note that the ISGMR strength distribution for $^{90}$Zr obtained in the present work is in excellent agreement with a previous measurement performed at RCNP \cite{Uchida_90Zr}. 

\begin{table*}
\caption{\label{tab:table1} Lorentzian-fit parameters and moment ratios for the ISGMR strength
distributions in $^{90,92}$Zr and $^{92}$Mo. All moment ratios are calculated over the $E_{x}$ range 10-30 MeV, where, $m_{k}=\int E^{k}_{x} S(E_{x}) dE_{x}$ is the $k^{th}$ moment of the strength distribution.}
\begin{tabular}{cccccccc}
\hline
  Nucleus     & $E_{m}$  (MeV)    & $\Gamma$    (MeV)  &EWSR & $m_{1}/m_{0}$ (MeV) &$\sqrt{m_{1}/m_{-1}}$ (MeV) &$\sqrt{m_{3}/m_{1}}$ (MeV) &$K_{A}$ (MeV)\\
\hline\\
$^{90}$Zr     &  16.55 $\pm$ 0.08   &  4.2 $\pm$ 0.3     & 0.95 $\pm$ 0.06 & 18.13 $\pm$ 0.09  & 17.66 $\pm$ 0.07 & 19.68 $\pm$ 0.13 &170.2 $\pm$ 2.2\\
$^{92}$Zr     &  16.12 $\pm$ 0.04   &  4.5 $\pm$ 0.2     & 0.97 $\pm$ 0.03 & 18.05 $\pm$ 0.05  & 17.52 $\pm$ 0.04 & 19.77 $\pm$ 0.06 &174.6 $\pm$ 1.1\\
$^{92}$Mo     &  16.79 $\pm$ 0.11   &  4.2 $\pm$ 0.4     & 0.84 $\pm$ 0.08 & 18.20 $\pm$ 0.13  & 17.76 $\pm$ 0.11 & 19.64 $\pm$ 0.21 &173.3 $\pm$ 3.8\\
\hline
\end{tabular}
\end{table*}

The ISGMR distributions were fitted with Lorentzian curves  and the associated peak energies ($E_{m}$)  and widths ($\Gamma$) for the three nuclei are nearly identical as well. A typical Lorentzian fit is also shown in Fig.~\ref{ISGMR}. The extracted parameters are presented in Table \ref{tab:table1}, along with the various moment ratios typically used in giant-resonance investigations, and nuclear incompressibility, $K_{A}$, determined from the moment ratio $\sqrt{m_{3}/m_{1}}$ which is employed in the scaling model for $E_\mathrm{ISGMR}$ \cite{str1982}. As is clear from the Table, the various moment ratios for all three nuclei, determined over the excitation energy range of 10 to 30 MeV, are also identical within experimental uncertainties,  as are the $K_A$ values obtained therefrom ($\sim$170 MeV in all cases). 

The EWSR fractions, determined over the 10--22 MeV excitation-energy range (encompassing the main ISGMR peak), are also provided in Table \ref{tab:table1}; in all cases, they  are close to 100\%. It should be noted that the quoted uncertainties in \%EWSR values are only statistical and do not include the systematic uncertainties (up to $\sim$20\%) arising from DWBA calculations, including those attributable to the choice of OM parameters (see, e.g., Ref. \cite{Li_2010}). 

Determination of the nuclear incompressibility from the ISGMR  is based on the assumption that the resonance energies do not depend on detailed structure of the nuclei involved. In fact, there had been no report prior to the results presented in Refs. \cite{YB_ZrMo2013, YB_Mo2015, Krishi_2015} of any  ``shell effects" leading to significant differences between ISGMR energies in nearby nuclei. For instance, measurements on three Lead isotopes,  $^{204,206,208}$Pb, had resulted in very similar ISGMR energies \cite{Darshna2013}. Further, detailed investigations of the ISGMR over the Sn and Cd isotopic chains have been performed in recent years \cite{Li_PRL2007, Li_2010, Darshana2012}. Although these nuclei emerged as ``soft" in comparison to the ``standard" nuclei, $^{90}$Zr and $^{208}$Pb, the ISGMR energies varied quite smoothly over a wide range of asymmetry parameter $(N-Z)/A$ \cite{Li_2010}. [This ``softness" of the nuclei in the Sn region is a separate question that has remained open till date \cite{fluffy1,fluffy2,fluffy6,fluffy5,fluffy4}.] Another structure effect appears due to deformation of nuclei, observed in the rare earth region \cite{garg_prl_1980, Itoh_prc2003, YB_154Sm_2004} and, more recently, in the nucleus $^{24}$Mg \cite{YKG_PLB2015}, which results in splitting of the strength keeping the ISGMR energy unchanged; the nuclei investigated in this work have no significant deformation, however. 

The present results, thus, establish clearly, and strongly, that determination of nuclear incompressibility in nearby medium-heavy to heavy nuclei is not influenced in any appreciable manner by the choice of specific nucleus, or by the underlying nuclear structure.

The obvious question is why the present results are so different from those obtained by the Texas A \& M group \cite{YB_ZrMo2013, YB_Mo2015, Krishi_2015}. We believe the answer lies in the way the ÒbackgroundÓ is accounted for in the two approaches. In the present work, all instrumental background is eliminated because of the superior optical properties of the Grand Raiden Spectrometer (see, e.g., Ref. \cite{Itoh_prc2003}, and 
Fig.~\ref{ZeroDeg}), leaving the physical continuum as part of the excitation-energy spectra. In the Texas A \& M work, an empirical background is subtracted by assuming that it has the shape of a straight line at high excitation, joining onto a Fermi shape at low excitation to model particle threshold effects \cite{dhybg,dhybg2}. This process subtracts the physical continuum as well. It is quite possible, and perhaps likely, that this background subtraction approach is responsible for the differing strengths observed for various nuclei in their work. Since there is no arbitrariness involved in the background-subtraction procedure employed in the present work, it may be argued that our final results are more reliable.

In summary, we have investigated the response of compressional ``breathing mode", the isoscalar giant monopole  resonance, in $^{90, 92}$Zr and $^{92}$Mo via inelastic scattering of 385-MeV $\alpha$ particles at extremely forward angles (including 0$^\circ$). The ISGMR response of these nuclei is practically identical, in contrast with recent reports where significant differences were observed in the ISGMR strength distributions for $^{92}$Zr and $^{92}$Mo as compared with that for $^{90}$Zr, claiming significant nuclear shell structure contributions to the nuclear incompressibility. 
The present results affirm the standard hydrodynamical picture associated with collective modes of oscillation and clearly indicate that the ISGMR strength distributions vary only in a minor way in nuclei near A$\sim$90. These variations, the origins of which are unclear, are too small to lead to any significant differences in the extracted nuclear incompressibilities, and it can be concluded that the incompressibilities are not affected by the shell structure of the nuclei near A$\sim$90.

The authors acknowledge the efforts of the staff of the RCNP Ring Cyclotron Facility in providing a high-quality, halo-free $\alpha$ beam required for the measurements reported here. This work has been supported in part by the U.S. National Science Foundation (Grant No. PHY-1419765).

\bibliography{UND_A90}

\begin{thebibliography}{42}
\expandafter\ifx\csname natexlab\endcsname\relax\def\natexlab#1{#1}\fi
\expandafter\ifx\csname bibnamefont\endcsname\relax
  \def\bibnamefont#1{#1}\fi
\expandafter\ifx\csname bibfnamefont\endcsname\relax
  \def\bibfnamefont#1{#1}\fi
\expandafter\ifx\csname citenamefont\endcsname\relax
  \def\citenamefont#1{#1}\fi
\expandafter\ifx\csname url\endcsname\relax
  \def\url#1{\texttt{#1}}\fi
\expandafter\ifx\csname urlprefix\endcsname\relax\def\urlprefix{URL }\fi
\providecommand{\bibinfo}[2]{#2}
\providecommand{\eprint}[2][]{\url{#2}}

\bibitem[{\citenamefont{Bohr and Mottelson}(1975)}]{BohrMott}
\bibinfo{author}{\bibfnamefont{A.}~\bibnamefont{Bohr}} \bibnamefont{and}
  \bibinfo{author}{\bibfnamefont{B.~R.} \bibnamefont{Mottelson}},
  \emph{\bibinfo{title}{Nuclear Structure}}, vol.~\bibinfo{volume}{II}
  (\bibinfo{publisher}{Benjamin, New York}, \bibinfo{year}{1975}).

\bibitem[{\citenamefont{Harakeh and van~der Woude}(1973)}]{Harakeh_book}
\bibinfo{author}{\bibfnamefont{M.~N.} \bibnamefont{Harakeh}} \bibnamefont{and}
  \bibinfo{author}{\bibfnamefont{A.}~\bibnamefont{van~der Woude}},
  \emph{\bibinfo{title}{Giant Resonances Fundamental High-Frequency Modes of
  Nuclear Excitation}} (\bibinfo{publisher}{Oxford University Press, New York,
  2001}, \bibinfo{year}{1973}).

\bibitem[{\citenamefont{Lattimer and Prakash}(2001)}]{Lattimer2001}
\bibinfo{author}{\bibfnamefont{J.~M.} \bibnamefont{Lattimer}} \bibnamefont{and}
  \bibinfo{author}{\bibfnamefont{M.}~\bibnamefont{Prakash}},
  \bibinfo{journal}{Astrophys. J.} \textbf{\bibinfo{volume}{550}},
  \bibinfo{pages}{426} (\bibinfo{year}{2001}).

\bibitem[{\citenamefont{Stringari}(1982)}]{str1982}
\bibinfo{author}{\bibfnamefont{S.}~\bibnamefont{Stringari}},
  \bibinfo{journal}{Phys. Lett. B} \textbf{\bibinfo{volume}{108}},
  \bibinfo{pages}{232} (\bibinfo{year}{1982}).

\bibitem[{\citenamefont{Treiner et~al.}(1981)\citenamefont{Treiner, Krivine,
  Bohigas, and Martorell}}]{Treiner1981}
\bibinfo{author}{\bibfnamefont{J.}~\bibnamefont{Treiner}},
  \bibinfo{author}{\bibfnamefont{H.}~\bibnamefont{Krivine}},
  \bibinfo{author}{\bibfnamefont{O.}~\bibnamefont{Bohigas}}, \bibnamefont{and}
  \bibinfo{author}{\bibfnamefont{J.}~\bibnamefont{Martorell}},
  \bibinfo{journal}{Nucl. Phys. A} \textbf{\bibinfo{volume}{371}},
  \bibinfo{pages}{253} (\bibinfo{year}{1981}).

\bibitem[{\citenamefont{Blaizot et~al.}(1995)\citenamefont{Blaizot, Berger,
  Decharg\'e, and Girod}}]{Blaizot1995}
\bibinfo{author}{\bibfnamefont{J.~P.} \bibnamefont{Blaizot}},
  \bibinfo{author}{\bibfnamefont{J.~F.} \bibnamefont{Berger}},
  \bibinfo{author}{\bibfnamefont{J.}~\bibnamefont{Decharg\'e}},
  \bibnamefont{and} \bibinfo{author}{\bibfnamefont{M.}~\bibnamefont{Girod}},
  \bibinfo{journal}{Nucl. Phys. A} \textbf{\bibinfo{volume}{591}},
  \bibinfo{pages}{435} (\bibinfo{year}{1995}).

\bibitem[{\citenamefont{Col\`o et~al.}(2004)\citenamefont{Col\`o, Giai, Meyer,
  Bennaceur, and Bonche}}]{Colo_Giai_PRC2004}
\bibinfo{author}{\bibfnamefont{G.}~\bibnamefont{Col\`o}},
  \bibinfo{author}{\bibfnamefont{N.~V.} \bibnamefont{Giai}},
  \bibinfo{author}{\bibfnamefont{J.}~\bibnamefont{Meyer}},
  \bibinfo{author}{\bibfnamefont{K.}~\bibnamefont{Bennaceur}},
  \bibnamefont{and} \bibinfo{author}{\bibfnamefont{P.}~\bibnamefont{Bonche}},
  \bibinfo{journal}{Phys. Rev. C} \textbf{\bibinfo{volume}{70}},
  \bibinfo{pages}{024307} (\bibinfo{year}{2004}).

\bibitem[{\citenamefont{Todd-Rutel and Piekarewicz}(2005)}]{Todd-Rutel_PRL2005}
\bibinfo{author}{\bibfnamefont{B.~G.} \bibnamefont{Todd-Rutel}}
  \bibnamefont{and}
  \bibinfo{author}{\bibfnamefont{J.}~\bibnamefont{Piekarewicz}},
  \bibinfo{journal}{Phys. Rev. Lett.} \textbf{\bibinfo{volume}{95}},
  \bibinfo{pages}{122501} (\bibinfo{year}{2005}).

\bibitem[{\citenamefont{Col\`o et~al.}(2014)\citenamefont{Col\`o, Garg, and
  Sagawa}}]{colo_garg_sagawa}
\bibinfo{author}{\bibfnamefont{G.}~\bibnamefont{Col\`o}},
  \bibinfo{author}{\bibfnamefont{U.}~\bibnamefont{Garg}}, \bibnamefont{and}
  \bibinfo{author}{\bibfnamefont{H.}~\bibnamefont{Sagawa}},
  \bibinfo{journal}{Eur. Phys. J. A} \textbf{\bibinfo{volume}{50}},
  \bibinfo{pages}{26} (\bibinfo{year}{2014}).

\bibitem[{\citenamefont{Piekarewicz}(2014)}]{pickarewickz2014}
\bibinfo{author}{\bibfnamefont{J.}~\bibnamefont{Piekarewicz}},
  \bibinfo{journal}{Eur. Phys. J. A} \textbf{\bibinfo{volume}{50}},
  \bibinfo{pages}{25} (\bibinfo{year}{2014}).

\bibitem[{\citenamefont{Youngblood et~al.}(2013)\citenamefont{Youngblood, Lui,
  Krishichayan, Button, Anders, Gorelik, Urin, and Shlomo}}]{YB_ZrMo2013}
\bibinfo{author}{\bibfnamefont{D.~H.} \bibnamefont{Youngblood}},
  \bibinfo{author}{\bibfnamefont{Y.-W.} \bibnamefont{Lui}},
  \bibinfo{author}{\bibnamefont{Krishichayan}},
  \bibinfo{author}{\bibfnamefont{J.}~\bibnamefont{Button}},
  \bibinfo{author}{\bibfnamefont{M.~R.} \bibnamefont{Anders}},
  \bibinfo{author}{\bibfnamefont{M.~L.} \bibnamefont{Gorelik}},
  \bibinfo{author}{\bibfnamefont{M.~H.} \bibnamefont{Urin}}, \bibnamefont{and}
  \bibinfo{author}{\bibfnamefont{S.}~\bibnamefont{Shlomo}},
  \bibinfo{journal}{Phys. Rev. C} \textbf{\bibinfo{volume}{88}},
  \bibinfo{pages}{021301} (\bibinfo{year}{2013}).

\bibitem[{\citenamefont{Youngblood et~al.}(2015)\citenamefont{Youngblood, Lui,
  Krishichayan, Button, Bonasera, and Shlomo}}]{YB_Mo2015}
\bibinfo{author}{\bibfnamefont{D.~H.} \bibnamefont{Youngblood}},
  \bibinfo{author}{\bibfnamefont{Y.-W.} \bibnamefont{Lui}},
  \bibinfo{author}{\bibnamefont{Krishichayan}},
  \bibinfo{author}{\bibfnamefont{J.}~\bibnamefont{Button}},
  \bibinfo{author}{\bibfnamefont{G.}~\bibnamefont{Bonasera}}, \bibnamefont{and}
  \bibinfo{author}{\bibfnamefont{S.}~\bibnamefont{Shlomo}},
  \bibinfo{journal}{Phys. Rev. C} \textbf{\bibinfo{volume}{92}},
  \bibinfo{pages}{014318} (\bibinfo{year}{2015}).

\bibitem[{\citenamefont{Krishichayan et~al.}(2015)\citenamefont{Krishichayan,
  Lui, Button, Youngblood, Bonasera, and Shlomo}}]{Krishi_2015}
\bibinfo{author}{\bibnamefont{Krishichayan}},
  \bibinfo{author}{\bibfnamefont{Y.-W.} \bibnamefont{Lui}},
  \bibinfo{author}{\bibfnamefont{J.}~\bibnamefont{Button}},
  \bibinfo{author}{\bibfnamefont{D.~H.} \bibnamefont{Youngblood}},
  \bibinfo{author}{\bibfnamefont{G.}~\bibnamefont{Bonasera}}, \bibnamefont{and}
  \bibinfo{author}{\bibfnamefont{S.}~\bibnamefont{Shlomo}},
  \bibinfo{journal}{Phys. Rev. C} \textbf{\bibinfo{volume}{92}},
  \bibinfo{pages}{044323} (\bibinfo{year}{2015}).

\bibitem[{\citenamefont{Harakeh et~al.}(1977)\citenamefont{Harakeh, van~der
  Borg, Ishimatsu, Morsch, and van~der Woude}}]{Harakesh_prl1977}
\bibinfo{author}{\bibfnamefont{M.~N.} \bibnamefont{Harakeh}},
  \bibinfo{author}{\bibfnamefont{K.}~\bibnamefont{van~der Borg}},
  \bibinfo{author}{\bibfnamefont{T.}~\bibnamefont{Ishimatsu}},
  \bibinfo{author}{\bibfnamefont{H.~P.} \bibnamefont{Morsch}},
  \bibnamefont{and} \bibinfo{author}{\bibfnamefont{A.}~\bibnamefont{van~der
  Woude}}, \bibinfo{journal}{Phys. Rev. Lett} \textbf{\bibinfo{volume}{38}},
  \bibinfo{pages}{676} (\bibinfo{year}{1977}).

\bibitem[{\citenamefont{Youngblood et~al.}(1977)\citenamefont{Youngblood, Roza,
  Moss, Brown, and Bronson}}]{YBPRL1977}
\bibinfo{author}{\bibfnamefont{D.~H.} \bibnamefont{Youngblood}},
  \bibinfo{author}{\bibfnamefont{C.~M.} \bibnamefont{Roza}},
  \bibinfo{author}{\bibfnamefont{J.~M.} \bibnamefont{Moss}},
  \bibinfo{author}{\bibfnamefont{D.~R.} \bibnamefont{Brown}}, \bibnamefont{and}
  \bibinfo{author}{\bibfnamefont{J.~D.} \bibnamefont{Bronson}},
  \bibinfo{journal}{Phys. Rev. Lett.} \textbf{\bibinfo{volume}{39}},
  \bibinfo{pages}{1188} (\bibinfo{year}{1977}).

\bibitem[{\citenamefont{Lipparini and Stringari}(1989)}]{LIPPARINI1989}
\bibinfo{author}{\bibfnamefont{E.}~\bibnamefont{Lipparini}} \bibnamefont{and}
  \bibinfo{author}{\bibfnamefont{S.}~\bibnamefont{Stringari}},
  \bibinfo{journal}{Phys. Rep.} \textbf{\bibinfo{volume}{175}},
  \bibinfo{pages}{103} (\bibinfo{year}{1989}).

\bibitem[{\citenamefont{Lattimer and Prakash}(2000)}]{lattpra}
\bibinfo{author}{\bibfnamefont{J.~M.} \bibnamefont{Lattimer}} \bibnamefont{and}
  \bibinfo{author}{\bibfnamefont{M.}~\bibnamefont{Prakash}},
  \bibinfo{journal}{Phys. Rep.} \textbf{\bibinfo{volume}{333-334}},
  \bibinfo{pages}{121} (\bibinfo{year}{2000}).

\bibitem[{\citenamefont{Fujiwara et~al.}(1999)\citenamefont{Fujiwara, Akimune,
  Daito, Fujimura, Fujita, Hatanaka, Ikegami, Katayama, Nagayama, Matsuoka
  et~al.}}]{Fujiwara_GR}
\bibinfo{author}{\bibfnamefont{M.}~\bibnamefont{Fujiwara}},
  \bibinfo{author}{\bibfnamefont{H.}~\bibnamefont{Akimune}},
  \bibinfo{author}{\bibfnamefont{I.}~\bibnamefont{Daito}},
  \bibinfo{author}{\bibfnamefont{H.}~\bibnamefont{Fujimura}},
  \bibinfo{author}{\bibfnamefont{Y.}~\bibnamefont{Fujita}},
  \bibinfo{author}{\bibfnamefont{K.}~\bibnamefont{Hatanaka}},
  \bibinfo{author}{\bibfnamefont{H.}~\bibnamefont{Ikegami}},
  \bibinfo{author}{\bibfnamefont{I.}~\bibnamefont{Katayama}},
  \bibinfo{author}{\bibfnamefont{K.}~\bibnamefont{Nagayama}},
  \bibinfo{author}{\bibfnamefont{N.}~\bibnamefont{Matsuoka}},
  \bibnamefont{et~al.}, \bibinfo{journal}{Nucl. Instrum. Meth. Phys. Res. A}
  \textbf{\bibinfo{volume}{422}}, \bibinfo{pages}{484} (\bibinfo{year}{1999}).

\bibitem[{\citenamefont{Itoh et~al.}(2003)\citenamefont{Itoh, Sakaguchi,
  Uchida, Ishikawa, Kawabata, Murakami, Takeda, Taki, Terashima, Tsukahara
  et~al.}}]{Itoh_prc2003}
\bibinfo{author}{\bibfnamefont{M.}~\bibnamefont{Itoh}},
  \bibinfo{author}{\bibfnamefont{H.}~\bibnamefont{Sakaguchi}},
  \bibinfo{author}{\bibfnamefont{M.}~\bibnamefont{Uchida}},
  \bibinfo{author}{\bibfnamefont{T.}~\bibnamefont{Ishikawa}},
  \bibinfo{author}{\bibfnamefont{T.}~\bibnamefont{Kawabata}},
  \bibinfo{author}{\bibfnamefont{T.}~\bibnamefont{Murakami}},
  \bibinfo{author}{\bibfnamefont{H.}~\bibnamefont{Takeda}},
  \bibinfo{author}{\bibfnamefont{T.}~\bibnamefont{Taki}},
  \bibinfo{author}{\bibfnamefont{S.}~\bibnamefont{Terashima}},
  \bibinfo{author}{\bibfnamefont{N.}~\bibnamefont{Tsukahara}},
  \bibnamefont{et~al.}, \bibinfo{journal}{Phys. Rev. C}
  \textbf{\bibinfo{volume}{68}}, \bibinfo{pages}{064602}
  (\bibinfo{year}{2003}).

\bibitem[{\citenamefont{Uchida et~al.}(2003)\citenamefont{Uchida, Sakaguchi,
  Itoh, Yosoi, Kawabata, Takeda, Yasuda, Murakami, Ishikawa, Taki
  et~al.}}]{Uchida_PLB2003}
\bibinfo{author}{\bibfnamefont{M.}~\bibnamefont{Uchida}},
  \bibinfo{author}{\bibfnamefont{H.}~\bibnamefont{Sakaguchi}},
  \bibinfo{author}{\bibfnamefont{M.}~\bibnamefont{Itoh}},
  \bibinfo{author}{\bibfnamefont{M.}~\bibnamefont{Yosoi}},
  \bibinfo{author}{\bibfnamefont{T.}~\bibnamefont{Kawabata}},
  \bibinfo{author}{\bibfnamefont{H.}~\bibnamefont{Takeda}},
  \bibinfo{author}{\bibfnamefont{Y.}~\bibnamefont{Yasuda}},
  \bibinfo{author}{\bibfnamefont{T.}~\bibnamefont{Murakami}},
  \bibinfo{author}{\bibfnamefont{T.}~\bibnamefont{Ishikawa}},
  \bibinfo{author}{\bibfnamefont{T.}~\bibnamefont{Taki}}, \bibnamefont{et~al.},
  \bibinfo{journal}{Phys. Lett. B} \textbf{\bibinfo{volume}{557}},
  \bibinfo{pages}{12} (\bibinfo{year}{2003}).

\bibitem[{\citenamefont{Gupta~et al.}()}]{yogesh2}
\bibinfo{author}{\bibfnamefont{Y.~K.} \bibnamefont{Gupta~et al.}},
  \bibinfo{howpublished}{\emph{to be published.}}

\bibitem[{\citenamefont{Brandenburg et~al.}(1987)\citenamefont{Brandenburg,
  Borghols, Drentje, Ekstr\"{o}m, Harakeh, van~der Woude, H\r{a}kanson,
  Nilsson, Olsson, Pignanelli et~al.}}]{BRANDENBURG1987}
\bibinfo{author}{\bibfnamefont{S.}~\bibnamefont{Brandenburg}},
  \bibinfo{author}{\bibfnamefont{W.~T.~A.} \bibnamefont{Borghols}},
  \bibinfo{author}{\bibfnamefont{A.~G.} \bibnamefont{Drentje}},
  \bibinfo{author}{\bibfnamefont{L.~P.} \bibnamefont{Ekstr\"{o}m}},
  \bibinfo{author}{\bibfnamefont{M.~N.} \bibnamefont{Harakeh}},
  \bibinfo{author}{\bibfnamefont{A.}~\bibnamefont{van~der Woude}},
  \bibinfo{author}{\bibfnamefont{A.}~\bibnamefont{H\r{a}kanson}},
  \bibinfo{author}{\bibfnamefont{L.}~\bibnamefont{Nilsson}},
  \bibinfo{author}{\bibfnamefont{N.}~\bibnamefont{Olsson}},
  \bibinfo{author}{\bibfnamefont{M.}~\bibnamefont{Pignanelli}},
  \bibnamefont{et~al.}, \bibinfo{journal}{Nucl. Phys. A}
  \textbf{\bibinfo{volume}{466}}, \bibinfo{pages}{29} (\bibinfo{year}{1987}).

\bibitem[{\citenamefont{Li et~al.}(2010)\citenamefont{Li, Garg, Liu, Marks,
  Nayak, \mbox{Madhusudhana} Rao, Fujiwara, Hashimoto, Nakanishi, Okumura
  et~al.}}]{Li_2010}
\bibinfo{author}{\bibfnamefont{T.}~\bibnamefont{Li}},
  \bibinfo{author}{\bibfnamefont{U.}~\bibnamefont{Garg}},
  \bibinfo{author}{\bibfnamefont{Y.}~\bibnamefont{Liu}},
  \bibinfo{author}{\bibfnamefont{R.}~\bibnamefont{Marks}},
  \bibinfo{author}{\bibfnamefont{B.~K.} \bibnamefont{Nayak}},
  \bibinfo{author}{\bibfnamefont{P.~V.} \bibnamefont{\mbox{Madhusudhana} Rao}},
  \bibinfo{author}{\bibfnamefont{M.}~\bibnamefont{Fujiwara}},
  \bibinfo{author}{\bibfnamefont{H.}~\bibnamefont{Hashimoto}},
  \bibinfo{author}{\bibfnamefont{K.}~\bibnamefont{Nakanishi}},
  \bibinfo{author}{\bibfnamefont{S.}~\bibnamefont{Okumura}},
  \bibnamefont{et~al.}, \bibinfo{journal}{Phys. Rev. C}
  \textbf{\bibinfo{volume}{81}}, \bibinfo{pages}{034309}
  (\bibinfo{year}{2010}).

\bibitem[{\citenamefont{Gupta et~al.}(2015)\citenamefont{Gupta, Garg, Matta,
  Patel, Peach, Hoffman, Yoshida, Itoh, Fujiwara, Hara et~al.}}]{YKG_PLB2015}
\bibinfo{author}{\bibfnamefont{Y.~K.} \bibnamefont{Gupta}},
  \bibinfo{author}{\bibfnamefont{U.}~\bibnamefont{Garg}},
  \bibinfo{author}{\bibfnamefont{J.}~\bibnamefont{Matta}},
  \bibinfo{author}{\bibfnamefont{D.}~\bibnamefont{Patel}},
  \bibinfo{author}{\bibfnamefont{T.}~\bibnamefont{Peach}},
  \bibinfo{author}{\bibfnamefont{J.}~\bibnamefont{Hoffman}},
  \bibinfo{author}{\bibfnamefont{K.}~\bibnamefont{Yoshida}},
  \bibinfo{author}{\bibfnamefont{M.}~\bibnamefont{Itoh}},
  \bibinfo{author}{\bibfnamefont{M.}~\bibnamefont{Fujiwara}},
  \bibinfo{author}{\bibfnamefont{K.}~\bibnamefont{Hara}}, \bibnamefont{et~al.},
  \bibinfo{journal}{Phys. Lett. B} \textbf{\bibinfo{volume}{748}},
  \bibinfo{pages}{343} (\bibinfo{year}{2015}).

\bibitem[{\citenamefont{Patel et~al.}(2012)\citenamefont{Patel, Garg, Fujiwara,
  Akimune, Berg, Harakeh, Itoh, Kawabata, Kawase, Nayak et~al.}}]{Darshana2012}
\bibinfo{author}{\bibfnamefont{D.}~\bibnamefont{Patel}},
  \bibinfo{author}{\bibfnamefont{U.}~\bibnamefont{Garg}},
  \bibinfo{author}{\bibfnamefont{M.}~\bibnamefont{Fujiwara}},
  \bibinfo{author}{\bibfnamefont{H.}~\bibnamefont{Akimune}},
  \bibinfo{author}{\bibfnamefont{G.~P.~A.} \bibnamefont{Berg}},
  \bibinfo{author}{\bibfnamefont{M.~N.} \bibnamefont{Harakeh}},
  \bibinfo{author}{\bibfnamefont{M.}~\bibnamefont{Itoh}},
  \bibinfo{author}{\bibfnamefont{T.}~\bibnamefont{Kawabata}},
  \bibinfo{author}{\bibfnamefont{K.}~\bibnamefont{Kawase}},
  \bibinfo{author}{\bibfnamefont{B.~K.} \bibnamefont{Nayak}},
  \bibnamefont{et~al.}, \bibinfo{journal}{Phys. Lett. B}
  \textbf{\bibinfo{volume}{718}}, \bibinfo{pages}{447} (\bibinfo{year}{2012}).

\bibitem[{\citenamefont{Satchler and Khoa}(1997)}]{Satchler_Khoa1997}
\bibinfo{author}{\bibfnamefont{G.~R.} \bibnamefont{Satchler}} \bibnamefont{and}
  \bibinfo{author}{\bibfnamefont{D.~T.} \bibnamefont{Khoa}},
  \bibinfo{journal}{Phys. Rev. C} \textbf{\bibinfo{volume}{55}},
  \bibinfo{pages}{285} (\bibinfo{year}{1997}).

\bibitem[{\citenamefont{Satchler}(1987)}]{Satchler1987}
\bibinfo{author}{\bibfnamefont{G.~R.} \bibnamefont{Satchler}},
  \bibinfo{journal}{Nucl. Phys. A} \textbf{\bibinfo{volume}{472}},
  \bibinfo{pages}{215} (\bibinfo{year}{1987}).

\bibitem[{\citenamefont{Harakeh and Dieperink}(1981)}]{Harakeh1981}
\bibinfo{author}{\bibfnamefont{M.~N.} \bibnamefont{Harakeh}} \bibnamefont{and}
  \bibinfo{author}{\bibfnamefont{A.~E.~L.} \bibnamefont{Dieperink}},
  \bibinfo{journal}{Phys. Rev. C} \textbf{\bibinfo{volume}{23}},
  \bibinfo{pages}{2329} (\bibinfo{year}{1981}).

\bibitem[{\citenamefont{Rhoades-Brown
  et~al.}(1980{\natexlab{a}})\citenamefont{Rhoades-Brown, Macfarlane, and
  Pieper}}]{ptolemy1}
\bibinfo{author}{\bibfnamefont{M.}~\bibnamefont{Rhoades-Brown}},
  \bibinfo{author}{\bibfnamefont{M.~H.} \bibnamefont{Macfarlane}},
  \bibnamefont{and} \bibinfo{author}{\bibfnamefont{S.~C.}
  \bibnamefont{Pieper}}, \bibinfo{journal}{Phys. Rev. C}
  \textbf{\bibinfo{volume}{21}}, \bibinfo{pages}{2417}
  (\bibinfo{year}{1980}{\natexlab{a}}).

\bibitem[{\citenamefont{Rhoades-Brown
  et~al.}(1980{\natexlab{b}})\citenamefont{Rhoades-Brown, Macfarlane, and
  Pieper}}]{ptolemy2}
\bibinfo{author}{\bibfnamefont{M.}~\bibnamefont{Rhoades-Brown}},
  \bibinfo{author}{\bibfnamefont{M.~H.} \bibnamefont{Macfarlane}},
  \bibnamefont{and} \bibinfo{author}{\bibfnamefont{S.~C.}
  \bibnamefont{Pieper}}, \bibinfo{journal}{Phys. Rev. C}
  \textbf{\bibinfo{volume}{21}}, \bibinfo{pages}{2436}
  (\bibinfo{year}{1980}{\natexlab{b}}).

\bibitem[{\citenamefont{Uchida et~al.}(2004)\citenamefont{Uchida, Sakaguchi,
  Itoh, Yosoi, Kawabata, Yasuda, Takeda, Murakami, Terashima, Kishi
  et~al.}}]{Uchida_90Zr}
\bibinfo{author}{\bibfnamefont{M.}~\bibnamefont{Uchida}},
  \bibinfo{author}{\bibfnamefont{H.}~\bibnamefont{Sakaguchi}},
  \bibinfo{author}{\bibfnamefont{M.}~\bibnamefont{Itoh}},
  \bibinfo{author}{\bibfnamefont{M.}~\bibnamefont{Yosoi}},
  \bibinfo{author}{\bibfnamefont{T.}~\bibnamefont{Kawabata}},
  \bibinfo{author}{\bibfnamefont{Y.}~\bibnamefont{Yasuda}},
  \bibinfo{author}{\bibfnamefont{H.}~\bibnamefont{Takeda}},
  \bibinfo{author}{\bibfnamefont{T.}~\bibnamefont{Murakami}},
  \bibinfo{author}{\bibfnamefont{S.}~\bibnamefont{Terashima}},
  \bibinfo{author}{\bibfnamefont{S.}~\bibnamefont{Kishi}},
  \bibnamefont{et~al.}, \bibinfo{journal}{Phys. Rev. C}
  \textbf{\bibinfo{volume}{69}}, \bibinfo{pages}{051301}
  (\bibinfo{year}{2004}).

\bibitem[{\citenamefont{Patel et~al.}(2013)\citenamefont{Patel, Garg, Fujiwara,
  Adachi, Akimune, Berg, Harakeh, Itoh, C.Iwamoto, Long et~al.}}]{Darshna2013}
\bibinfo{author}{\bibfnamefont{D.}~\bibnamefont{Patel}},
  \bibinfo{author}{\bibfnamefont{U.}~\bibnamefont{Garg}},
  \bibinfo{author}{\bibfnamefont{M.}~\bibnamefont{Fujiwara}},
  \bibinfo{author}{\bibfnamefont{T.}~\bibnamefont{Adachi}},
  \bibinfo{author}{\bibfnamefont{H.}~\bibnamefont{Akimune}},
  \bibinfo{author}{\bibfnamefont{G.~P.~A.} \bibnamefont{Berg}},
  \bibinfo{author}{\bibfnamefont{M.~N.} \bibnamefont{Harakeh}},
  \bibinfo{author}{\bibfnamefont{M.}~\bibnamefont{Itoh}},
  \bibinfo{author}{\bibnamefont{C.Iwamoto}},
  \bibinfo{author}{\bibfnamefont{A.}~\bibnamefont{Long}}, \bibnamefont{et~al.},
  \bibinfo{journal}{Phys. Lett. B} \textbf{\bibinfo{volume}{726}},
  \bibinfo{pages}{178} (\bibinfo{year}{2013}).

\bibitem[{\citenamefont{Li et~al.}(2007)\citenamefont{Li, Garg, Liu, Marks,
  Nayak, \mbox{Madhusudhana} Rao, Fujiwara, Hashimoto, Kawase, Nakanishi
  et~al.}}]{Li_PRL2007}
\bibinfo{author}{\bibfnamefont{T.}~\bibnamefont{Li}},
  \bibinfo{author}{\bibfnamefont{U.}~\bibnamefont{Garg}},
  \bibinfo{author}{\bibfnamefont{Y.}~\bibnamefont{Liu}},
  \bibinfo{author}{\bibfnamefont{R.}~\bibnamefont{Marks}},
  \bibinfo{author}{\bibfnamefont{B.~K.} \bibnamefont{Nayak}},
  \bibinfo{author}{\bibfnamefont{P.~V.} \bibnamefont{\mbox{Madhusudhana} Rao}},
  \bibinfo{author}{\bibfnamefont{M.}~\bibnamefont{Fujiwara}},
  \bibinfo{author}{\bibfnamefont{H.}~\bibnamefont{Hashimoto}},
  \bibinfo{author}{\bibfnamefont{K.}~\bibnamefont{Kawase}},
  \bibinfo{author}{\bibfnamefont{K.}~\bibnamefont{Nakanishi}},
  \bibnamefont{et~al.}, \bibinfo{journal}{Phys. Rev. Lett.}
  \textbf{\bibinfo{volume}{99}}, \bibinfo{pages}{162503}
  (\bibinfo{year}{2007}).

\bibitem[{\citenamefont{Garg et~al.}(2007)\citenamefont{Garg, Li, Okumura,
  Akimune, Fujiwara, Harakeh, Hashimoto, Itoh, Iwao, Kawabata
  et~al.}}]{fluffy1}
\bibinfo{author}{\bibfnamefont{U.}~\bibnamefont{Garg}},
  \bibinfo{author}{\bibfnamefont{T.}~\bibnamefont{Li}},
  \bibinfo{author}{\bibfnamefont{S.}~\bibnamefont{Okumura}},
  \bibinfo{author}{\bibfnamefont{H.}~\bibnamefont{Akimune}},
  \bibinfo{author}{\bibfnamefont{M.}~\bibnamefont{Fujiwara}},
  \bibinfo{author}{\bibfnamefont{M.}~\bibnamefont{Harakeh}},
  \bibinfo{author}{\bibfnamefont{H.}~\bibnamefont{Hashimoto}},
  \bibinfo{author}{\bibfnamefont{M.}~\bibnamefont{Itoh}},
  \bibinfo{author}{\bibfnamefont{Y.}~\bibnamefont{Iwao}},
  \bibinfo{author}{\bibfnamefont{T.}~\bibnamefont{Kawabata}},
  \bibnamefont{et~al.}, \bibinfo{journal}{Nucl. Phys. A}
  \textbf{\bibinfo{volume}{788}}, \bibinfo{pages}{36} (\bibinfo{year}{2007}).

\bibitem[{\citenamefont{Piekarewicz}(2007)}]{fluffy2}
\bibinfo{author}{\bibfnamefont{J.}~\bibnamefont{Piekarewicz}},
  \bibinfo{journal}{Phys. Rev. C} \textbf{\bibinfo{volume}{76}},
  \bibinfo{pages}{031301} (\bibinfo{year}{2007}).

\bibitem[{\citenamefont{Piekarewicz}(2010)}]{fluffy6}
\bibinfo{author}{\bibfnamefont{J.}~\bibnamefont{Piekarewicz}},
  \bibinfo{journal}{J. Phys. G: Nucl. Part. Phys.}
  \textbf{\bibinfo{volume}{37}}, \bibinfo{pages}{064038}
  (\bibinfo{year}{2010}).

\bibitem[{\citenamefont{\mbox{Li-Gang Cao}
  et~al.}(2012)\citenamefont{\mbox{Li-Gang Cao}, Sagawa, and Col\`o}}]{fluffy5}
\bibinfo{author}{\bibnamefont{\mbox{Li-Gang Cao}}},
  \bibinfo{author}{\bibfnamefont{H.}~\bibnamefont{Sagawa}}, \bibnamefont{and}
  \bibinfo{author}{\bibfnamefont{G.}~\bibnamefont{Col\`o}},
  \bibinfo{journal}{Phys. Rev. C} \textbf{\bibinfo{volume}{86}},
  \bibinfo{pages}{054313} (\bibinfo{year}{2012}).

\bibitem[{\citenamefont{Vesel\'y et~al.}(2012)\citenamefont{Vesel\'y, Toivanen,
  Carlsson, Dobaczewski, Michel, and Pastore}}]{fluffy4}
\bibinfo{author}{\bibfnamefont{P.}~\bibnamefont{Vesel\'y}},
  \bibinfo{author}{\bibfnamefont{J.}~\bibnamefont{Toivanen}},
  \bibinfo{author}{\bibfnamefont{B.~G.} \bibnamefont{Carlsson}},
  \bibinfo{author}{\bibfnamefont{J.}~\bibnamefont{Dobaczewski}},
  \bibinfo{author}{\bibfnamefont{N.}~\bibnamefont{Michel}}, \bibnamefont{and}
  \bibinfo{author}{\bibfnamefont{A.}~\bibnamefont{Pastore}},
  \bibinfo{journal}{Phys. Rev. C} \textbf{\bibinfo{volume}{86}},
  \bibinfo{pages}{024303} (\bibinfo{year}{2012}).

\bibitem[{\citenamefont{Garg et~al.}(1980)\citenamefont{Garg, Bogucki, Bronson,
  Lui, Rozsa, and Youngblood}}]{garg_prl_1980}
\bibinfo{author}{\bibfnamefont{U.}~\bibnamefont{Garg}},
  \bibinfo{author}{\bibfnamefont{P.}~\bibnamefont{Bogucki}},
  \bibinfo{author}{\bibfnamefont{J.~D.} \bibnamefont{Bronson}},
  \bibinfo{author}{\bibfnamefont{Y.-W.} \bibnamefont{Lui}},
  \bibinfo{author}{\bibfnamefont{C.~M.} \bibnamefont{Rozsa}}, \bibnamefont{and}
  \bibinfo{author}{\bibfnamefont{D.~H.} \bibnamefont{Youngblood}},
  \bibinfo{journal}{Phys. Rev. Lett.} \textbf{\bibinfo{volume}{45}},
  \bibinfo{pages}{1670} (\bibinfo{year}{1980}).

\bibitem[{\citenamefont{Youngblood
  et~al.}(2004{\natexlab{a}})\citenamefont{Youngblood, Lui, Clark, John,
  Tokimoto, and Chen}}]{YB_154Sm_2004}
\bibinfo{author}{\bibfnamefont{D.~H.} \bibnamefont{Youngblood}},
  \bibinfo{author}{\bibfnamefont{Y.-W.} \bibnamefont{Lui}},
  \bibinfo{author}{\bibfnamefont{H.~L.} \bibnamefont{Clark}},
  \bibinfo{author}{\bibfnamefont{B.}~\bibnamefont{John}},
  \bibinfo{author}{\bibfnamefont{Y.}~\bibnamefont{Tokimoto}}, \bibnamefont{and}
  \bibinfo{author}{\bibfnamefont{X.}~\bibnamefont{Chen}},
  \bibinfo{journal}{Phys. Rev. C} \textbf{\bibinfo{volume}{69}},
  \bibinfo{pages}{034315} (\bibinfo{year}{2004}{\natexlab{a}}).

\bibitem[{\citenamefont{Youngblood
  et~al.}(2004{\natexlab{b}})\citenamefont{Youngblood, Lui, Clark, John,
  Tokimoto, and Chen}}]{dhybg}
\bibinfo{author}{\bibfnamefont{D.~H.} \bibnamefont{Youngblood}},
  \bibinfo{author}{\bibfnamefont{Y.-W.} \bibnamefont{Lui}},
  \bibinfo{author}{\bibfnamefont{H.~L.} \bibnamefont{Clark}},
  \bibinfo{author}{\bibfnamefont{B.}~\bibnamefont{John}},
  \bibinfo{author}{\bibfnamefont{Y.}~\bibnamefont{Tokimoto}}, \bibnamefont{and}
  \bibinfo{author}{\bibfnamefont{X.}~\bibnamefont{Chen}},
  \bibinfo{journal}{Phys. Rev. C} \textbf{\bibinfo{volume}{69}},
  \bibinfo{pages}{034315} (\bibinfo{year}{2004}{\natexlab{b}}).

\bibitem[{\citenamefont{Youngblood}()}]{dhybg2}
\bibinfo{author}{\bibfnamefont{D.~H.} \bibnamefont{Youngblood}},
  \bibinfo{note}{\em private communication}.

\end{thebibliography}

\end{document}